\documentclass[
reprint,
amsmath,
amssymb,
aps,
prd,
tightenlines
]{revtex4-1}

\usepackage{graphicx}
\usepackage{xcolor}
\usepackage{siunitx}
\usepackage{booktabs}
\usepackage{multirow}
\usepackage{overpic}
\usepackage{colortbl}
\usepackage{array}
\usepackage{makecell}
\usepackage{subfigure}
\usepackage{dcolumn}
\usepackage{bm}
\usepackage[normalem]{ulem}
\usepackage{enumitem}

\definecolor{aa}{RGB}{0,0,139}

\setlist[enumerate,1]{
  label=(\Roman*),
  leftmargin=*,
  nosep
}

\setlist[enumerate,2]{
  label=(\roman*),
  leftmargin=1.5em,
  nosep
}

\definecolor{boslv}{rgb}{0.0, 0.65, 0.58}
\definecolor{Munsell}{HTML}{00A877}

\newcommand{\psipp}{\psi(3686)}

\newcommand{\bfg}{\begin{figure}}
\newcommand{\efg}{\end{figure}}
\newcommand{\bitm}{\begin{itemize}}
\newcommand{\eitm}{\end{itemize}}
\newcommand{\bnum}{\begin{enumerate}}
\newcommand{\enum}{\end{enumerate}}
\newcommand{\btbl}{\begin{table}}
\newcommand{\etbl}{\end{table}}
\newcommand{\btbu}{\begin{tabular}}
\newcommand{\etbu}{\end{tabular}}
\newcommand{\bcl}{\begin{center}}
\newcommand{\ecl}{\end{center}}

\newcommand{\beq}{\begin{equation}}
\newcommand{\eeq}{\end{equation}}
\newcommand{\beqr}{\begin{eqnarray}}
\newcommand{\eeqr}{\end{eqnarray}}

\usepackage[
  colorlinks=true,
  urlcolor=blue,
  citecolor=blue,
  linkcolor=blue
]{hyperref}

\newcommand{\BESIIIorcid}[1]{%
  \href{https://orcid.org/#1}{%
    \hspace*{0.1em}%
    \raisebox{-0.45ex}{%
      \includegraphics[width=1em]{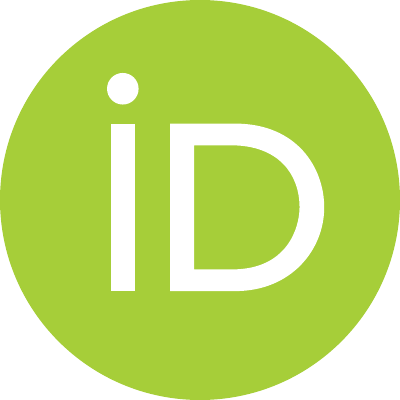}%
    }%
  }%
}

\begin{document}

\title{\texorpdfstring{{\boldmath Observation of $\eta_{c} \to p\bar{p}\eta$ via $\psi(3686) \to \gamma p\bar{p}\eta$}}{Observation of eta-c to p anti-p eta via psi(3686) to gamma p anti-p eta}}

\author{
\begin{center}
M.~Ablikim$^{1}$\BESIIIorcid{0000-0002-3935-619X},
M.~N.~Achasov$^{4,c}$\BESIIIorcid{0000-0002-9400-8622},
P.~Adlarson$^{83}$\BESIIIorcid{0000-0001-6280-3851},
X.~C.~Ai$^{89}$\BESIIIorcid{0000-0003-3856-2415},
C.~S.~Akondi$^{31A,31B}$\BESIIIorcid{0000-0001-6303-5217},
R.~Aliberti$^{39}$\BESIIIorcid{0000-0003-3500-4012},
A.~Amoroso$^{82A,82C}$\BESIIIorcid{0000-0002-3095-8610},
Q.~An$^{78,65,\dagger}$,
Y.~H.~An$^{89}$\BESIIIorcid{0009-0008-3419-0849},
M.~S.~Anderson$^{39}$\BESIIIorcid{0009-0008-1550-2632}
% ...
\\
\vspace{0.2cm}
(BESIII Collaboration)\\
\vspace{0.2cm} {\it
$^{1}$ Institute of High Energy Physics, Beijing 100049, People's Republic of China\\
% ...
$^{4}$ Budker Institute of Nuclear Physics SB RAS (BINP), Novosibirsk 630090, Russia\\
% ...
$^{31}$ INFN Sezione di Ferrara, (A)INFN Sezione di Ferrara, I-44122, Ferrara, Italy; (B)University of Ferrara, I-44122, Ferrara, Italy\\
% ...
$^{39}$ Johannes Gutenberg University of Mainz, Johann-Joachim-Becher-Weg 45, D-55099 Mainz, Germany\\
% ...
$^{65}$ State Key Laboratory of Particle Detection and Electronics, Beijing 100049, Hefei 230026, People's Republic of China\\
% ...
$^{78}$ University of Science and Technology of China, Hefei 230026, People's Republic of China\\
% ...
$^{82}$ University of Turin and INFN, (A)University of Turin, I-10125, Turin, Italy; (B)University of Eastern Piedmont, I-15121, Alessandria, Italy; (C)INFN, I-10125, Turin, Italy\\
$^{83}$ Uppsala University, Box 516, SE-75120 Uppsala, Sweden\\
% ...
$^{89}$ Zhengzhou University, Zhengzhou 450001, People's Republic of China\\

\vspace{0.2cm}
$^{\dagger}$ Deceased\\
% ...
$^{c}$ Also at the Novosibirsk State University, Novosibirsk, 630090, Russia\\
% ...

}

\end{center}
\vspace{0.4cm}
}

\begin{abstract}
The decay $\eta_c\to p\bar{p}\eta$ is observed for the first time with a significance of exceeding $10\sigma$. It is found by analyzing $(2712.4 \pm 14.3)\times10^{6}$ $\psi(3686)$ events accumulated at the BESIII detector. The measured branching fraction of $\eta_c\to p\bar{p}\eta$ via $\psi(3686) \to \gamma p \bar{p} \eta$ is significantly influenced by the interference between the resonant $\eta_c$ decay and the non-resonant process $\psi(3686) \to \gamma p \bar{p} \eta$ and is measured in both constructive- and destructive-interference scenarios. 
The joint branching fraction of $\psi(3686)\to \gamma\eta_c$, $\eta_c\to p\bar{p}\eta$ is measured to be $(3.2 \pm 0.1 \pm 0.9)\times10^{-6}$ or $(8.7 \pm 0.3 \pm 2.1)\times10^{-6}$ for constructive- or destructive-interference solutions, respectively, where the first uncertainties are statistical and the second systematic. The branching fraction of $\eta_c\to p\bar{p}\eta$ is determined to be $\mathcal{B}(\eta_c\to p\bar{p}\eta)=(0.90 \pm 0.04 \pm 0.21 \pm 0.13)\times10^{-3}$ or $(2.42 \pm 0.07 \pm 0.48 \pm 0.34)\times10^{-3}$ for the two solutions, respectively, where the third uncertainties are due to the uncertainty in the branching fraction of $\psi(3686)\to \gamma \eta_c$. 
\end{abstract}


\maketitle

\section{INTRODUCTION}
Charmonium states below the open-charm production threshold have been extensively studied. The spin singlets~\cite{spin_singlets}, such as the $P$-wave state $h_c$, the $S$-wave ground state $\eta_c$, and its first radial excitation $\eta_c(2S)$, remain less well understood, particularly with regard to their decay properties. Up to now, the known decay modes of $\eta_c$ account for only about $70\%$ of its total branching fraction (BF), and many decay modes remain unknown. In the known decay modes of $\eta_c$, the BF for $\eta_c$ decays into final states involving baryons, such as $p\bar{p}$, $p\bar{p}\pi^0$, $\Lambda\bar\Lambda$, $\bar\Lambda(1520)\Lambda + c.c.$, $\Sigma^+\bar\Sigma^-$, $\Xi^-\bar\Xi^+$, $K^+\bar{p}\Lambda + c.c.$, and $\pi^+\pi^-p\bar{p}$, is at the level of $2\%$, according to the Particle Data Group (PDG)~\cite{ParticleDataGroup:2024cfk}. For $\eta_c$ decays into baryon–antibaryon pairs accompanied by a pseudoscalar meson ($\mathrm{B\bar{B}M}$), only two channels have been reported: $\eta_c \to p\bar{p}\pi^0$ observed by BESIII~\cite{BESIII:2012urf}, and $\eta_c \to K^+\Lambda\bar{p} + c.c.$ observed by Belle~\cite{Belle:2018ies}. Searching for new $\mathrm{B\bar{B}M}$ decay modes of the $\eta_c$ is essential to complete the experimental inventory of the $\eta_c$ decays and to investigate the dynamics of baryon-antibaryon pair production.

The limited knowledge of $\eta_c$ decays stems from the inability to produce the $\eta_{c}$ meson directly in $e^+e^-$ collisions. Therefore, most measurements of $\eta_c$ decays rely on the magnetic dipole ($M\it1$) transitions from $J/\psi$ or hindered $M\it1$ transitions from $\psi(3686)$, in addition to the electric dipole ($E\it1$) transition $h_c\to\gamma\eta_c$ via $\psi(3686)\to\pi^0 h_c$~\cite{guoaq} and $e^+e^-\to\pi^+\pi^-h_c$~\cite{maxn} as well as the hindered electromagnetic Dalitz (EMD) decay $\psi(3686) \to e^+ e^- \eta_c$~\cite{miaohan}. Although the interference between the $\eta_c$ and non-$\eta_c$ amplitudes in hindered $M1$ transitions distorts the $\eta_c$ line shape more significantly than it does in the $E1$ or hindered EMD channels~\cite{etac}, these $M1$ transitions nevertheless provide an important channel for searching for new $\eta_c$ decays. This is due to their higher $\eta_c$ production yields and lower background levels. The latter benefit results from the radiative photon's relatively high energy, which effectively suppresses combinatorial backgrounds.

Recently, BESIII observed a significant signal of the process $h_c\to p \bar{p} \eta$ with a significance of 5.1$\sigma$~\cite{hc}, and the BF of $h_c\to p \bar{p} \eta$ is reported to be $(7.4\pm2.2)\times10^{-4}$ in PDG~\cite{ParticleDataGroup:2024cfk}.
Based on perturbative Quantum Chromodynamics (pQCD) and non-relativistic Quantum Chromodynamics (NRQCD), the ratio of the total hadronic decay widths between $h_c$ and $\eta_c$,
$\Gamma_{h_c}^{\mathrm{hadronic}}/\Gamma_{\eta_c}^{\mathrm{hadronic}}$, is predicted to be $0.010\pm0.001$ by pQCD or $0.083\pm0.018$ by NRQCD~\cite{Y_P_Kuang}. Applying these predictions on the corresponding exclusive decay modes, the BF for the process $\eta_c\to p\bar{p}\eta$ is expected to be $(1.89\pm0.59)\times10^{-3}$ by pQCD or $(2.28\pm0.84)\times10^{-4}$ by NRQCD. 
The world's largest $\psi(3686)$ data sample of BESIII~\cite{Ablikim:2017wyh,liucheng} makes this first measurement of $\eta_c \to p \bar{p} \eta$ accessible.

We report the first observation of the decay $\eta_c\to p\bar{p}\eta$ via the $M1$ transition $\psi(3686)\rightarrow\gamma\eta_c$. The result is based on a sample of $(2712.4\pm14.3)\times10^6$ $\psi(3686)$ events~\cite{Ablikim:2017wyh,liucheng} collected with the BESIII detector. The analysis includes a treatment of the interference between $\eta_c$ and non-$\eta_c$ signal components.
The BF of $\eta_c\to p \bar{p}\eta$ is presented and compared with the theoretical calculations.

\section{DETECTOR AND DATA SAMPLES}
\label{sec:BES}

The BESIII detector~\cite{Ablikim:2009aa} records symmetric $e^+e^-$ collisions 
provided by the BEPCII storage ring~\cite{Yu:IPAC2016-TUYA01} in the center-of-mass energy range from 1.84 to 4.95~GeV, with a peak luminosity of $1.1 \times 10^{33}\;\text{cm}^{-2}\text{s}^{-1}$ achieved at $\sqrt{s} = 3.773\;\text{GeV}$. BESIII has collected large data samples in this energy region~\cite{BESIII:2020nme,Lu:2020imt,Zhang:2022bdc}. The cylindrical core of the BESIII detector covers $93\%$ of the full solid angle and consists of a helium-based multilayer drift chamber~(MDC), a time-of-flight~(TOF) system, and a CsI(Tl) electromagnetic calorimeter~(EMC), which are all enclosed in a superconducting solenoidal magnet providing a 1.0~T magnetic field. The solenoid is supported by an octagonal flux-return yoke with resistive plate counter muon identification modules interleaved with steel. 
The charged-particle momentum resolution at $1~{\rm GeV}/c$ is $0.5\%$, and the 
${\rm d}E/{\rm d}x$ resolution is $6\%$ for electrons from Bhabha scattering. The EMC measures photon energies with a resolution of $2.5\%$ ($5\%$) at $1$~GeV in the barrel (end cap) region. The time resolution in the plastic scintillator TOF barrel region is 68~ps, while that in the end cap region was 110~ps. The end cap TOF system was upgraded in 2015 using multigap resistive plate chamber
technology, providing a time resolution of 60~ps, which benefits $83\%$ of the data used in this analysis~\cite{Li:2017jpg,Guo:2017sjt,Cao:2020ibk}.

Monte Carlo (MC) simulated samples produced with a {\sc geant4}-based~\cite{geant4} software package, which includes the geometric description~\cite{Huang2022} of the BESIII detector and its response, are used to determine detection efficiencies and to estimate backgrounds. The
simulation models the beam energy spread and initial state radiation (ISR) in the $e^+e^-$ annihilations with the generator {\sc kkmc}~\cite{kkmc_a,kkmc_b}. The inclusive MC sample includes the production of the $\psi(3686)$ resonance, the ISR production of the $J/\psi$, and the continuum processes incorporated in {\sc kkmc}~\cite{kkmc_a,kkmc_b}. Particle decays are modelled with {\sc evtgen}~\cite{evtgen_a,evtgen_b} using BFs either taken from the PDG~\cite{ParticleDataGroup:2024cfk} when available, or otherwise estimated with {\sc lundcharm}~\cite{lundcharm_a,lundcharm_b}. Final state radiation from charged final state particles is incorporated using the
{\sc photos} package~\cite{photos}.

To study the detection efficiencies, signal MC sample of $\psi(3686)\to\gamma\eta_c\to\gamma p\bar{p}\eta$ with $\eta\to\gamma\gamma$ is generated, with an angular distribution of $(1+ \rm{\cos^2\theta_{\gamma}})$ for $\psi(3686)\to \gamma \eta_c$, where $\theta_{\gamma}$ is the angle between the photon and the positron beam direction in the center-of-mass system. 

To accurately simulate the dynamics of the possible intermediate states for $\eta_c\to p\bar{p}\eta$, additional MC samples are also generated.

\section{EVENT SELECTION}
\label{sec:selection}

Charged tracks detected in the MDC are required to be within a polar angle ($\theta$) range of $|\rm{\cos\theta}|<0.93$, where $\theta$ is measured from the $z$-axis, which is the symmetry axis of the MDC. The distance of the closest approach to the interaction point (IP) must be less than 10\,cm
along the $z$-axis, and less than 1\,cm in the transverse plane.  A candidate event must have two good charged tracks with opposite charges. Particle identification~(PID) for charged tracks combines measurements of the energy deposited in the MDC~(d$E$/d$x$) and the flight time in the TOF to form likelihoods $\mathcal{L}(h)~(h=p,K,\pi)$ for each hadron $h$ hypothesis. Tracks are identified as protons when the proton hypothesis has the greatest likelihood ($\mathcal{L}(p)>\mathcal{L}(K)$, $\mathcal{L}(p)>\mathcal{L}(\pi)$ and $\mathcal{L}(p)>0.001$). Finally, the events containing one proton and one anti-proton are retained.

Photon candidates are identified using isolated showers in the EMC. The deposited energy of each shower must be greater than 25 MeV in the barrel region ($|\cos\theta|<0.80$) and more than 50 MeV in the end cap regions ($0.86<|\cos\theta|<0.92$). To suppress electronic noise and showers from beam background, the difference between the EMC time and the event start time is required to be within [0,700] $\mathrm{ns}$. 
To suppress fake photons, the angle between the photon and the extrapolated impact point in the EMC of the nearest proton (anti-proton) track must be larger than 10 (20) degrees. The number of photon candidates must be at least 3 per event.

In order to suppress the remaining backgrounds and to improve the mass resolution, a four-momentum conservation constraint (4C) kinematic fit under the hypothesis of $\psi(3686)\to3\gamma p\bar{p}$ is performed. Furthermore, the requirement $\chi_{\mathrm{4C}}^{2}<65$ is imposed, determined through an optimization of the figure-of-merit (FOM), defined as ${\rm{FOM}} = \frac{S}{\sqrt{S+B}}$~\cite{punzi}. Here, $S$ denotes the number of signal MC events and is normalized to the data sample, based on the preliminary result of the constructive-interference solution. The $B$ denotes the number of events from the inclusive MC sample excluding the signal channel, and has been normalized to the size of the data. 
Moreover, to suppress backgrounds with two or four photons, additional 4C kinematic fits are performed for both signal and background channels by looping over all photon combinations according to different hypotheses of photon numbers. The one with the least $\chi^2$ is kept for each channel. We then require that $\chi^{2}_{\mathrm{4C}}(3\gamma p\bar{p})$ to be less than both $\chi^{2}_{\mathrm{4C}}(2\gamma p\bar{p})$ and $\chi^{2}_{\mathrm{4C}}(4\gamma p\bar{p})$.

\begin{figure*}[htbp]
    \centering
    \begin{minipage}{0.329\textwidth}
        \centering
        \includegraphics[width=\linewidth]{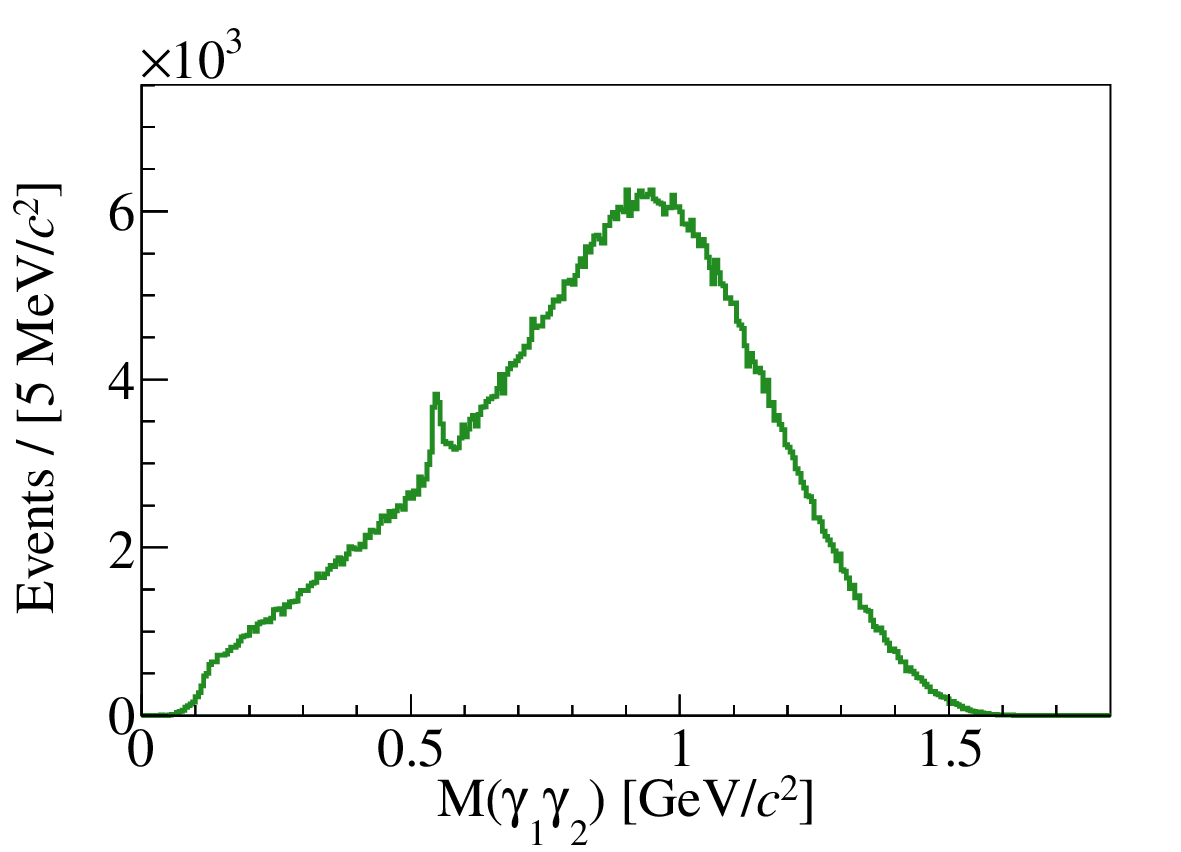}
        \put(-30,90){\fontsize{7}{14}\selectfont\color{black}(a)}
    \end{minipage}
    \hfill
    \begin{minipage}{0.329\textwidth}
        \centering
        \includegraphics[width=\linewidth]{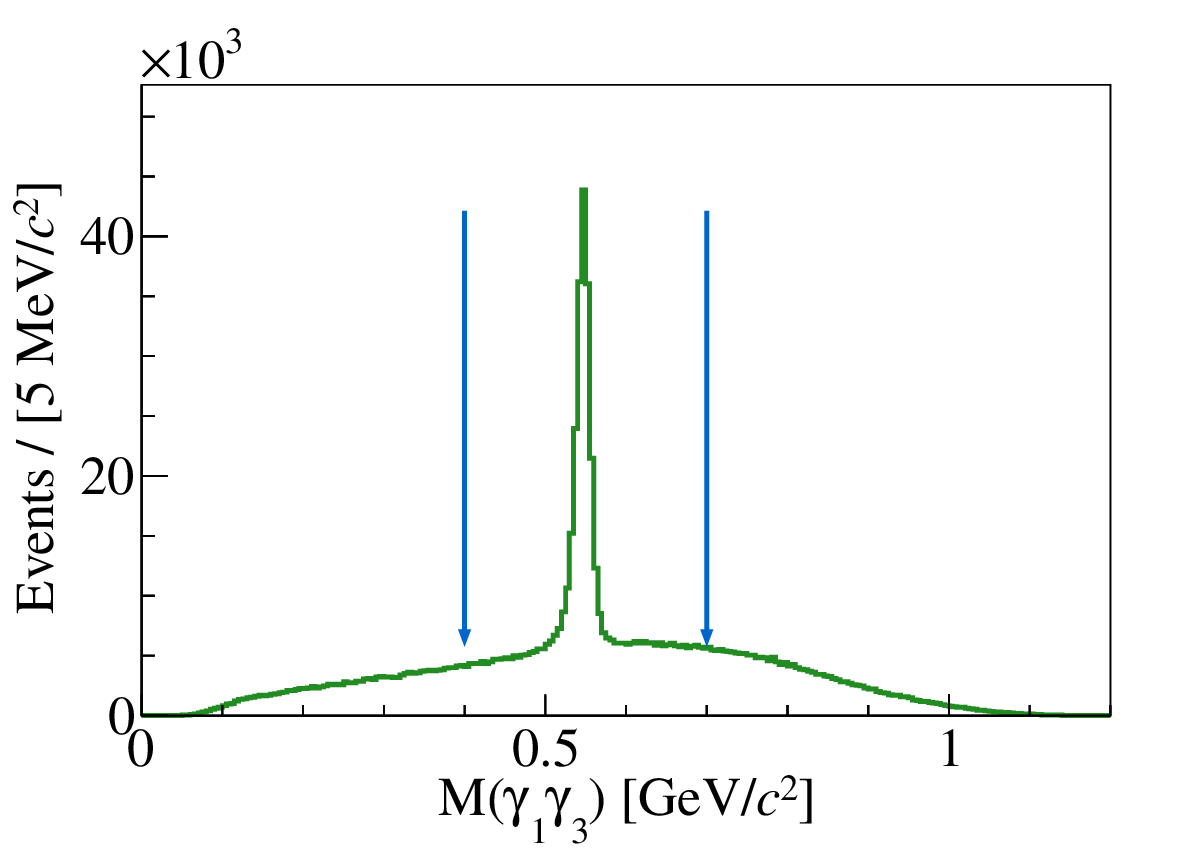}
        \put(-30,90){\fontsize{7}{14}\selectfont\color{black}(b)}
    \end{minipage}
    \hfill
    \begin{minipage}{0.329\textwidth}
        \centering
        \includegraphics[width=\linewidth]{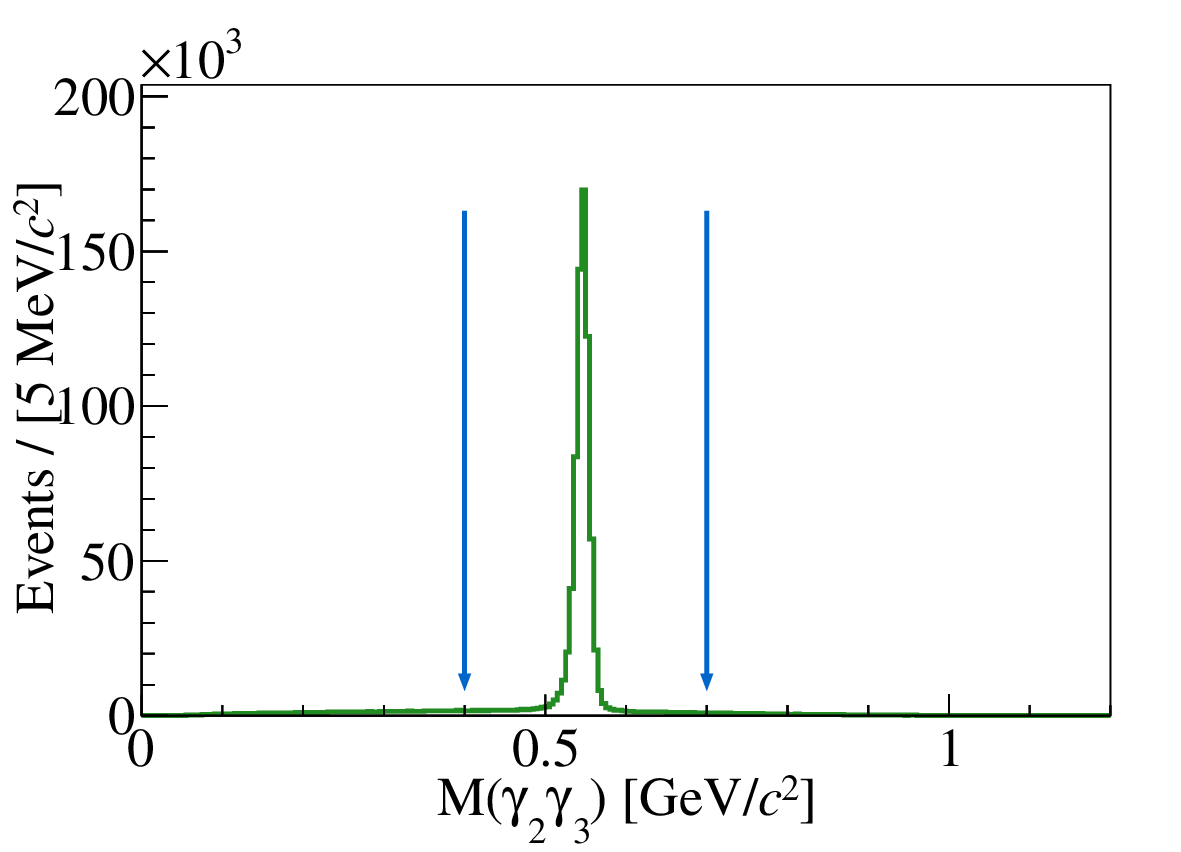}
        \put(-30,90){\fontsize{7}{14}\selectfont\color{black}(c)}
    \end{minipage}
    \caption{Invariant mass distributions of (a) $\gamma_{1}\gamma_{2}$, (b) $\gamma_{1}\gamma_{3}$, and (c) $\gamma_{2}\gamma_{3}$ from the signal MC sample (before $\eta$ reconstruction). The blue arrows indicate the range of the (0.4, 0.7)~$\mathrm{GeV}/c^{2}$ $\eta$ mass cut.}
    \label{fig:hist_m2g_sigMC_before_eta_rec}
\end{figure*}

After the 4C kinematic fit, the $3\gamma p \bar{p}$ hypothesis candidate is selected. For each selected candidate, the photons are labeled according to their energy in the laboratory frame, with $\gamma_1$ having the highest energy, $\gamma_2$ the next highest, and $\gamma_3$ the lowest. Based on the invariant mass distributions of these photon pairs from the signal MC sample (shown in Fig.~\ref{fig:hist_m2g_sigMC_before_eta_rec}), we apply the following sequential conditions in the $\eta$ meson reconstruction: (1) If the invariant mass $M(\gamma_2\gamma_3)$ falls within the $\eta$ mass window $(0.4, 0.7) ~\mathrm{GeV}/c^2$, the photon pair $(\gamma_2, \gamma_3)$ is assigned to the $\eta$ decay. (2) Otherwise, if $M(\gamma_1\gamma_3)$ falls within the same $\eta$ mass window $(0.4, 0.7) ~\mathrm{GeV}/c^2$, the photon pair $(\gamma_1, \gamma_3)$ is selected for the $\eta$ reconstruction. (3) If neither condition is satisfied, the event is rejected.

To suppress $\pi^0\to\gamma\gamma$ events, the invariant mass of any two-photon pair must fall outside the $\pi^0$ mass windows. The left and right boundaries of the mass windows are optimized simultaneously using the same method as that employed for $\chi^2_{4C}$, i.e., by optimizing a FOM. The optimized windows for the three $\gamma\gamma$ combinations are very similar and are all around the nominal $\pi^0$ mass, approximately corresponding to $M(\gamma\gamma)\notin(0.12,0.15)\,\mathrm{GeV}/c^2$.

To suppress the background of $\psi(3686)\to\eta J/\psi$ events, the recoiling mass of $\gamma_2 \gamma_3$, $RM(\gamma_2 \gamma_3)$, must fall outside the $J/\psi$ mass window.
Here the boundaries of the mass window are optimized by using the same method as for vetoing $\pi^0$, and they are determined to be $RM(\gamma_2 \gamma_3)\notin(3.072,3.125)~\mathrm{GeV}/c^2$.

\begin{figure}[htbp]
    \begin{minipage}{0.45\textwidth}
    \centering
    \includegraphics[width=1\textwidth]{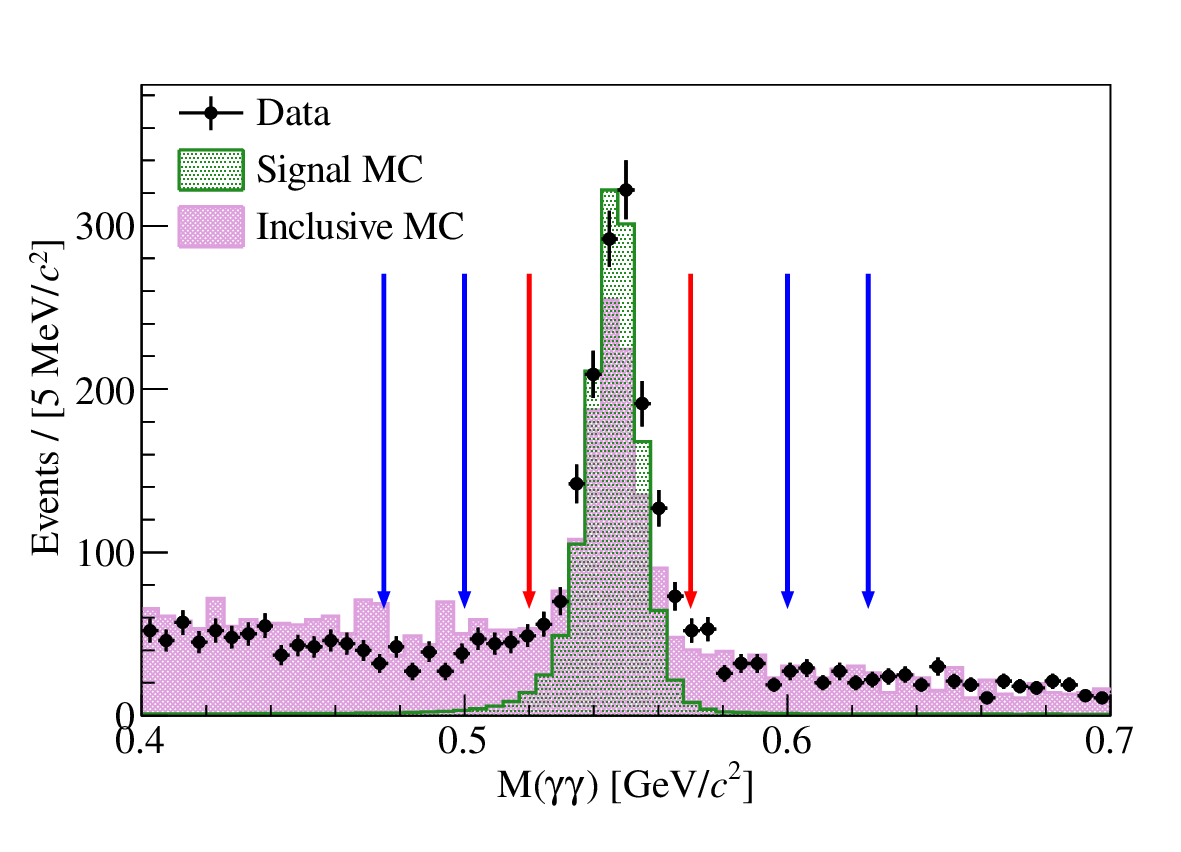}
    \end{minipage}
    \caption{The invariant mass distribution of the photon pair that is reconstructed as an $\eta$ meson after applying all the selection criteria. The black dots with error bars represent the data. The green and pink histograms represent the signal MC sample and the inclusive MC sample, respectively. The inclusive MC sample is normalized to the area of the data sample, and the signal MC sample is normalized to the bin with the highest number of data events. The red arrows indicate the signal range $(0.520, 0.570)~\mathrm{GeV}/c^{2}$, and the blue arrows indicate the sideband regions $(0.475, 0.500)$ or $(0.600, 0.625)~\mathrm{GeV}/c^{2}$.}
    \label{fig:hist_meta}
\end{figure}

After applying all the selection criteria, the invariant mass distribution of this photon pair is shown in Fig.~\ref{fig:hist_meta}. To avoid bias, the signal region for the $\eta$ meson is chosen to be approximately $2.5\sigma$, corresponding to $(0.520, 0.570)~\mathrm{GeV}/c^2$, while the side band regions are determined to be within $(0.475, 0.500)$ or $(0.600, 0.625)~\mathrm{GeV}/c^2$.

The remaining potential backgrounds are investigated with the inclusive $\psi(3686)$ MC samples, using the event-type analysis tool {\sc topoana}~\cite{zhouxy_topoAna}. The results show that there is no peaking background contribution in the $M(p \bar{p} \eta)$ spectrum. The remaining background events are dominated by $\psi(3686)\to\gamma \chi_{cJ}(J=1,2)$, $\chi_{cJ}\to\gamma J/\psi$, $J/\psi\to p\bar{p}\eta$, and $\psi(3686)\to\pi^0 J/\psi$, $J/\psi\to p\bar{p}\eta$.

Backgrounds from the continuum process $e^+e^-\to \gamma^{*} \to p\bar{p} \eta$ are investigated using the data samples collected at a center-of-mass energy of $3.65~\mathrm{GeV}$ with an integrated luminosity of 454 $\rm{pb}^{-1}$~\cite{cont}. After applying all the selection criteria and requiring candidates to fall in the $\eta$ and $\eta_c$ signal regions, two events are observed in the $3.65~\mathrm{GeV}$ continuum data sample. Taking into account the normalization of the integrated luminosity and the center-of-mass energy dependence of the cross section, this corresponds to only about eighteen events in the $\psi(3686)$ data sample. Therefore, the non-interference backgrounds are modeled using the inclusive MC simulation without introducing a specific continuum background component.

\section{SIGNAL YIELD and BRANCHING FRACTION}
\label{Sec:BR_determined}
The $\eta_c$ signal yield is obtained by performing an unbinned maximum likelihood fit to the $M(p \bar{p} \eta)$ distribution. Interference between the $\eta_c$ and the non-resonant process is considered. Interference between the $J/\psi$ and the non-resonant background is not considered, because the symmetric line shape of the $J/\psi$ resonance indicates that any interfering amplitude between the resonant signal and the non-resonant background is negligible. The probability density function (PDF) of the $\eta_c$ signal is expressed as 
\begin{equation}
\label{pdf}
\begin{aligned}
   &\Bigg(\epsilon(m) \Bigg| e^{i \phi} \sqrt{E_\gamma^7 f_d\left(E_\gamma\right)} B W_1(m) + \sqrt{f_{bg_1}} \Bigg|^2 \Bigg) \otimes G\\
   &+ N_{\mathrm{non}}\times f_{bg_2} + \left|B W_2(m)\right|^2 \otimes G.
\end{aligned}
\end{equation}
Here, $m$ denotes $M(p \bar{p} \eta)$; $\epsilon(m)$ is the efficiency function parameterized as a  $2^{\rm{nd}}$-order Chebychev function determined by the MC simulation; $E_{\gamma}=\frac{m_{\psi(2S)}^2 - m^2}{2 m_{\psi(2S)}}$ is the energy of the $M\it1$ transition photon; and  $E^7_\gamma$ is the radiative photon factor representing the hindered $M\it1$ transition  $\psi(3686)\to \gamma \eta_{c}$; $f_d(E_\gamma)=\frac{E_{0}^2}{E_{\gamma}E_{0}+(E_{\gamma}-E_{0})^2}$ is the damping function used by the KEDR experiment~\cite{KEDR} to damp the divergent tail raised by $E_{\gamma}^7$, where $E_{0}=\frac{m_{\psi(2S)}^2 - m_{\eta_c}^2}{2 m_{\psi(2S)}}$ is the nominal energy of the transition photon; \mbox{$BW_{j}(m)=\frac{1}{m^2-m_0^2+im_0\Gamma}$} is the Breit-Wigner function, where $m_0$ and $\Gamma$ are free parameters for $\eta_c$ when $j = 1$, and $m_0$ and $\Gamma$ are fixed to the world averaged values~\cite{ParticleDataGroup:2024cfk} for $J/\psi$ when $j = 2$. The non-resonant process interfering with $\eta_c$, denoted by $f_{bg_1}$, is described by a $2^{\rm nd}$-order polynomial function with floating parameters. The interference phase angle $\phi$ is a free parameter in the fit. The Gaussian function $G$, with floating parameters, accounts for the detector resolution. The PDF for non-interference backgrounds, denoted by $f_{bg_2}$, is modeled using the shape of the inclusive MC sample where backgrounds with a $\gamma p\bar{p}\eta$ final state or those containing $J/\psi \to p\bar{p}\eta$ decays have been removed. The yield of this component, $N_{non}$, is normalized to the integrated luminosity of the data sample and is subsequently fixed in the fit procedure.

\begin{figure*}[!t]
\centering
\includegraphics[width=0.45\textwidth]{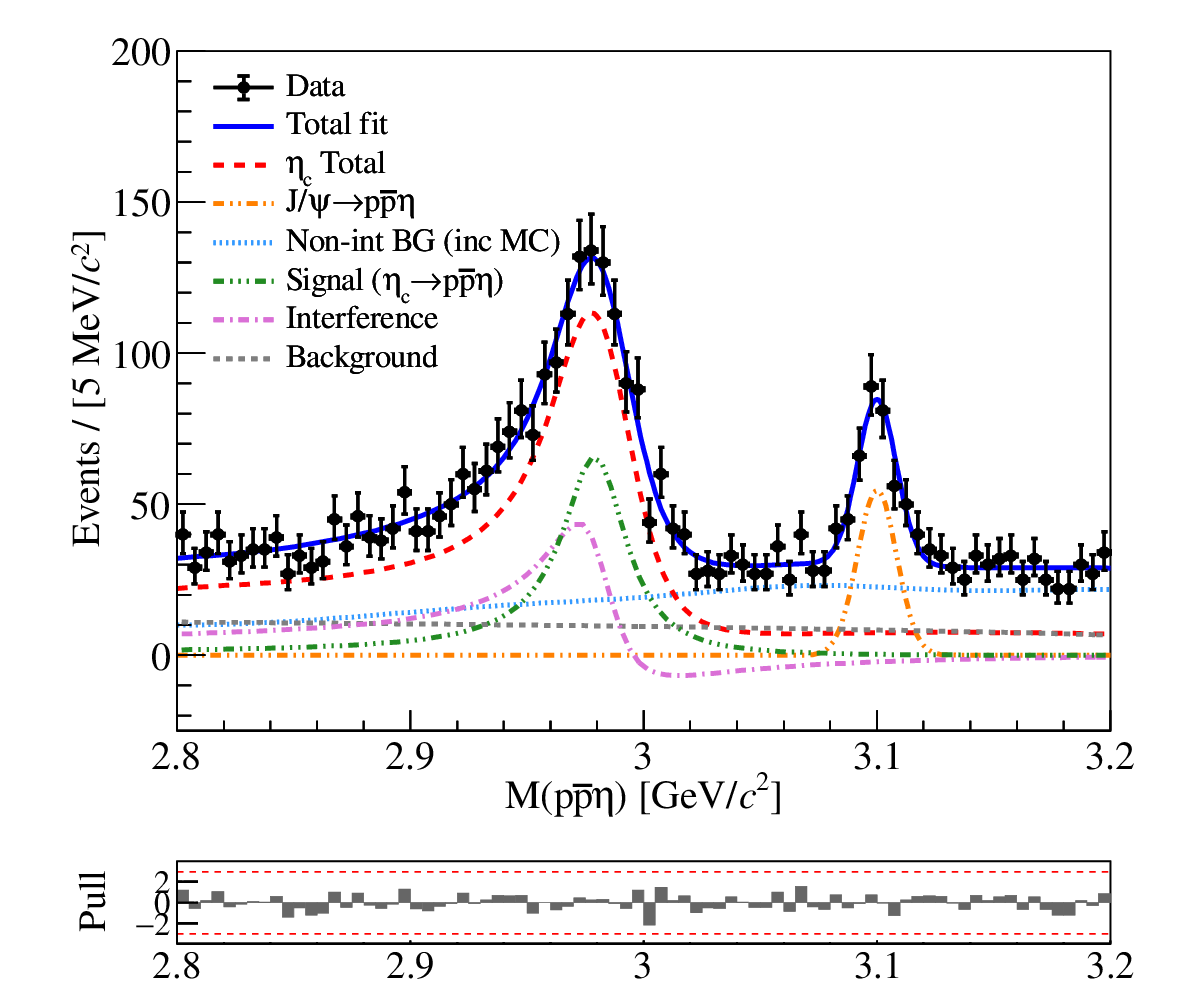}
\put(-50,170){\fontsize{9}{14}\selectfont\color{black}(a)}
\hfill
\includegraphics[width=0.45\textwidth]{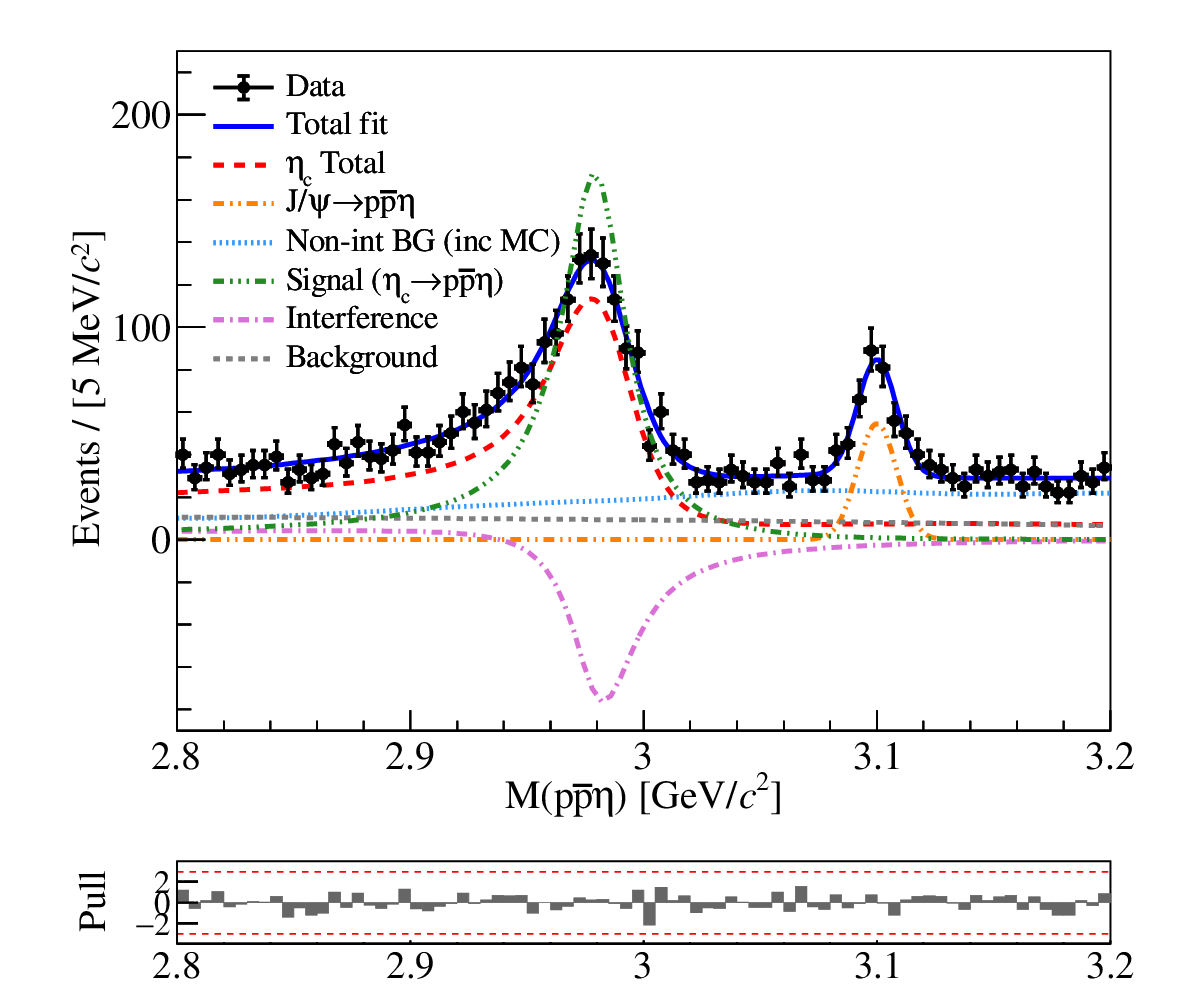}
\put(-50,170){\fontsize{9}{14}\selectfont\color{black}(b)}
\caption{The $M(p\bar{p}\eta)$ spectrum in data and the related fit results of (a)constructive interference and (b) destructive interference. The goodness of the fit is $\chi^2/ndf = 45/69$ in both cases. Black dots with error bars show data. The solid blue curve represents the total fit. The dashed red curve indicates the total shape of $\eta_c \to p \bar{p} \eta$. The dash-dotted orange curve shows the signal shape of $J / \psi \to p \bar{p} \eta$, the dotted light-blue curve represents the non-interference background from inclusive MC sample where backgrounds with a $\gamma p\bar{p}\eta$ final state or those containing $J/\psi \to p\bar{p}\eta$ decays have been removed, the dotted green curve is the signal shape of $\eta_c \to p \bar{p} \eta$, the dotted purple curve shows the interference, and the dotted gray curve shows the non-resonant process.}
\label{fig:fitting_result_metac}
\end{figure*}

According to the calculation in Ref.~\cite{mul_solution}, there should be two solutions for the fit to the $M(p \bar{p} \eta)$ distribution with the PDF in Eq.(\ref{pdf}). The interference phase angle is scanned to check for multiple solutions, where all the parameters are free except the interference phase angle. In the scan, the interference phase is changed from $0~\mathrm{rad}$ to $2\pi ~\mathrm{rad}$ in steps of $0.04~\mathrm{rad}$. Two minimal values are found at $\phi = 2.40~\mathrm{rad}$ (constructive) and $\phi =4.32~\mathrm{rad}$ (destructive). These two values are used only as initial values for the nominal fits. The fitted phase angles, $\phi=2.39\pm0.15~\mathrm{rad}$ (constructive) and $\phi=4.34\pm0.09~\mathrm{rad}$ (destructive), are consistent with the minima found in the scan, indicating that the scan correctly identified the two interference solutions. The fit results are shown in Fig.~\ref{fig:fitting_result_metac}. The goodness of fit is $\chi^2/ndf = 45/69$, and is exactly the same for both solutions. The statistical significance of $\eta_c\to p\bar{p}\eta$ is determined to be larger than $10\sigma$.

\begin{figure*}[htbp]
    \centering
    \begin{minipage}{0.329\textwidth}
        \centering
        \includegraphics[width=\linewidth]{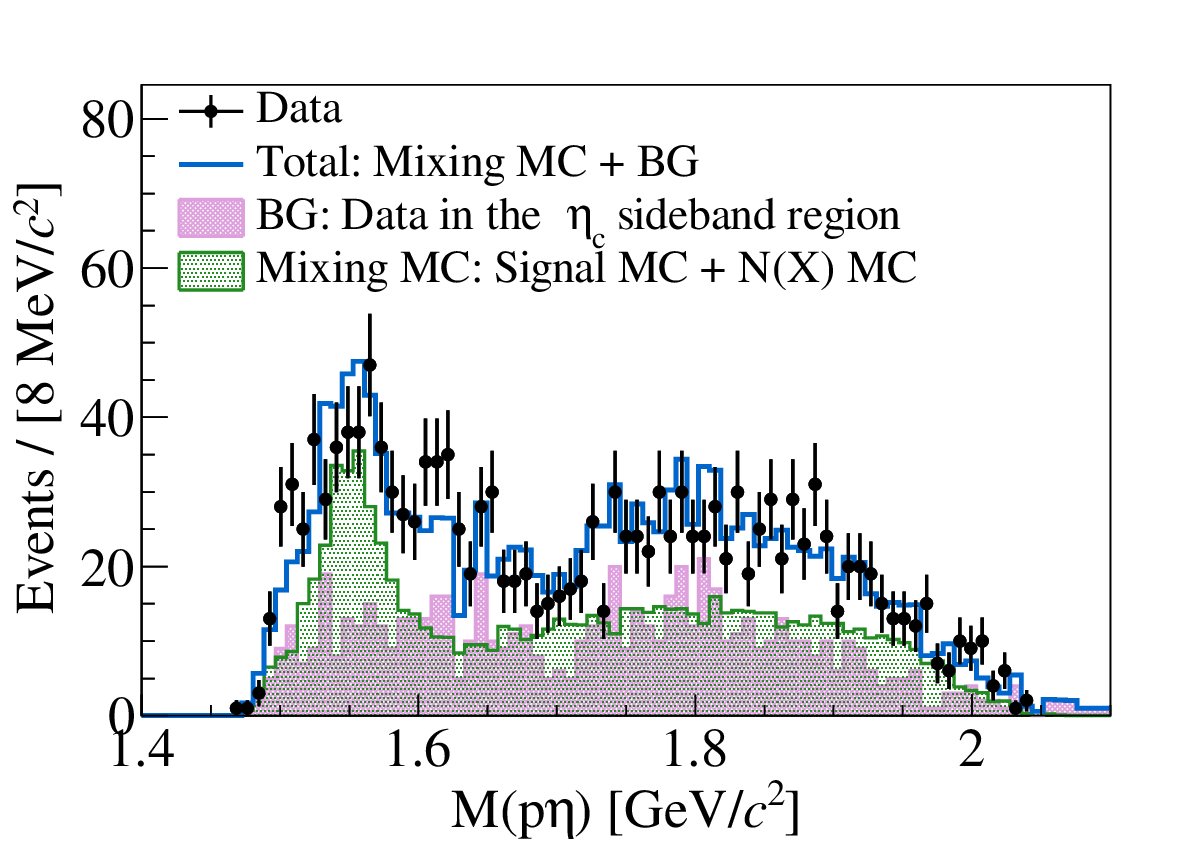}
        \put(-30,95){\fontsize{7}{14}\selectfont\color{black}(a)}
    \end{minipage}
    \hfill
    \begin{minipage}{0.329\textwidth}
        \centering
        \includegraphics[width=\linewidth]{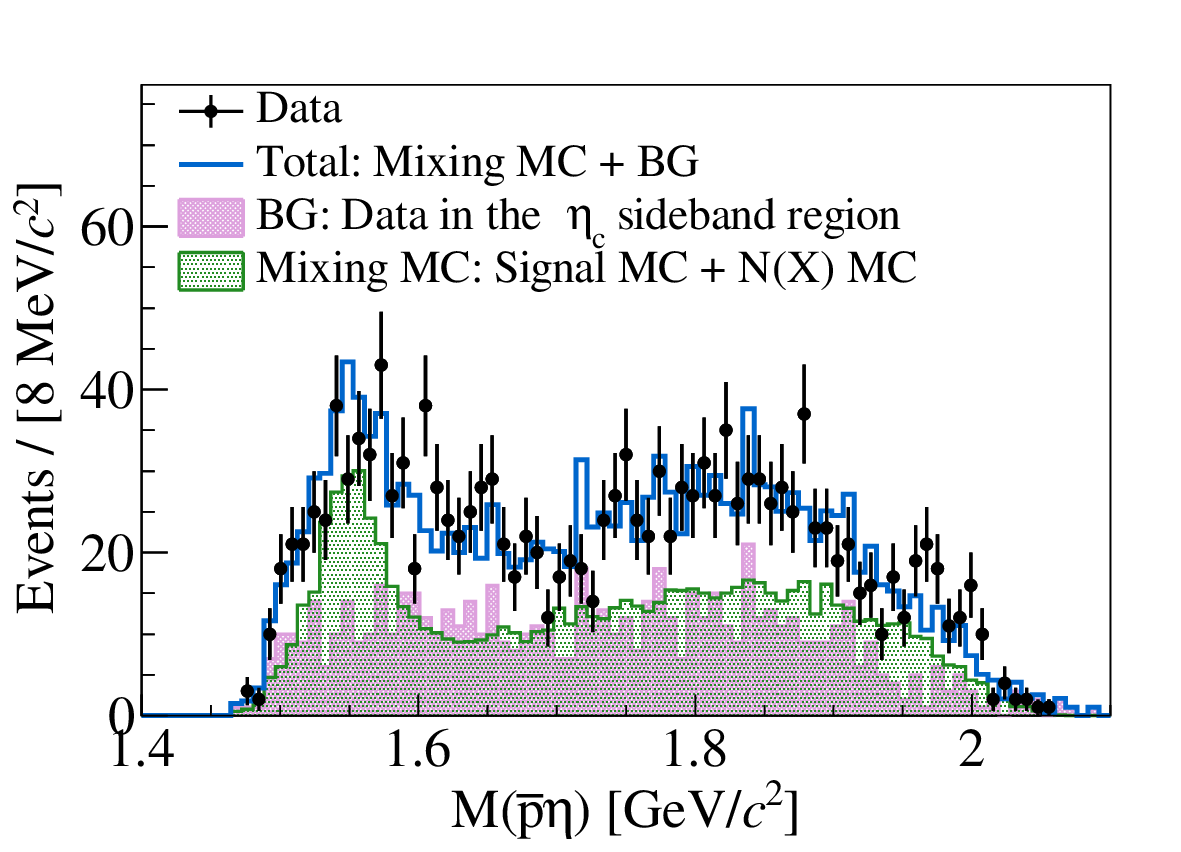}
        \put(-30,95){\fontsize{7}{14}\selectfont\color{black}(b)}
    \end{minipage}
    \hfill
    \begin{minipage}{0.329\textwidth}
        \centering
        \includegraphics[width=\linewidth]{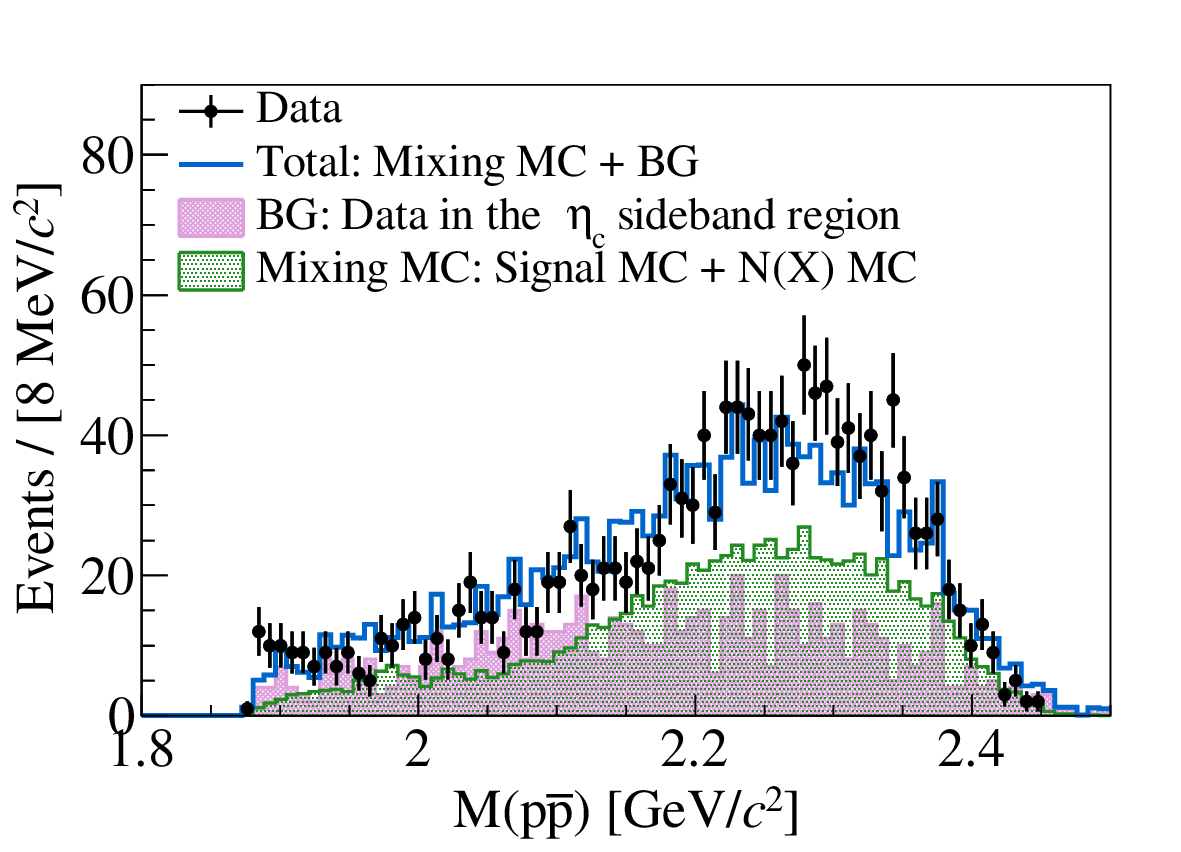}
        \put(-30,95){\fontsize{7}{14}\selectfont\color{black}(c)}
    \end{minipage}
    \caption{Two-body invariant mass distributions of the selected $\eta_{c} \to p \bar{p} \eta$ candidates: (a) $M(p\eta)$, (b) $M(\bar{p}\eta)$, and (c) $M(p\bar{p})$. The black dots with error bars denote data, the blue histograms denote the mixing MC simulation plus the background contribution, the pink histograms denote the background contribution described by the data in the $\eta_{c}$ side band regions and the green histograms denote the nominal mixing MC simulation.}
    \label{fig:mpeta_mpbeta_mppb_mixMC}
\end{figure*}

Potential intermediate states are investigated in the $p\bar{p}$, $p\eta$ and $\bar{p}\eta$ invariant mass spectra as shown in Fig.~\ref{fig:mpeta_mpbeta_mppb_mixMC}. The possible excited baryon $N(1535)$ ($\bar{N}(1535)$) can be seen in the mass distribution of $p\eta$ ($\bar{p}\eta$), but no significant $p\bar{p}$ threshold enhancement is observed. To accurately simulate the dynamics of the possible intermediate state for $\eta_c\to p\bar{p}\eta$ decay, two methods are adopted to generate MC samples. The first one is a mixing of a $N(1535)$ baryon and phase space (PHSP), where the mass and width of the $N(1535)$ baryon are fixed to the world averaged values~\cite{ParticleDataGroup:2024cfk} and the ratio between the two is determined by fitting the $M(p/\bar{p}\eta)$ distributions. The second one is a mixing of a $N(X)$ resonance and PHSP, where the mass and width of the $N(X)$ resonance are determined to be $1.554 ~\mathrm{GeV}/c^2$ and $0.043~\mathrm{GeV}$ by fitting to the $M(p/\bar{p}\eta)$ distributions in data. 
The fit incorporates three components: the shape from the PHSP MC sample in the $\eta_c$ signal region, a Breit–Wigner function representing the $N(X)$ resonance, and the shape from the data sample in the $\eta_c$ sideband region.  
The fractions of the three components are also determined by fitting the $M(p/\bar{p}\eta)$ distributions. Comparison between the MC simulations and data indicates that the data can be described better using the second mixing MC method, as shown in Fig.~\ref{fig:mpeta_mpbeta_mppb_mixMC}. Therefore, the second mixing MC sample is used to estimate the nominal detector efficiency, and the efficiency difference between the two mixing MC samples is taken as the uncertainty of the physics model.

The BF for $\eta_c \to p \bar{p} \eta$ is calculated by
\begin{align}
\mathcal{B}(\eta_c\to p\bar{p}\eta)=\frac{N^{\mathrm{obs}}}{N_{\psi(3686)}^{\mathrm{tot}}\times\mathcal{B}_{1}\times\mathcal{B}_{2}\times\epsilon},
\end{align}
where $N^{\mathrm{obs}}$ is the number of signal events obtained by fitting, $N_{\psi(3686)}^{\mathrm{tot}}$ is the total number of $\psi(3686)$ events, $\mathcal{B}_{1}$ and $\mathcal{B}_{2}$ are the BFs of $\psi(3686)\to \gamma \eta_c$ and $\eta\to \gamma \gamma$ taken from PDG~\cite{ParticleDataGroup:2024cfk}, respectively, and $\epsilon$ is the detection efficiency after correcting the track helix parameters in the MC simulation as described in Sec.~\ref{sec:sysU}.

Table~\ref{table:fit_metappb} lists the number of $\eta_c$ signal events, BFs, and other relevant values.

\begin{table}[htbp]
\begin{center}
\caption{The obtained mass, width, phase angle, signal yields, and BF of the decay $\eta_c \to p \bar{p}\eta$ for the two interference solutions. The uncertainties are statistical only.}
\label{table:fit_metappb} 
\begin{tabular}{l c c}
\hline\hline 
Solution & Constructive & Destructive \\
\hline 
$M(\eta_c)$[MeV/$c^2]$ & $2980.2 \pm 2.0$ & $2980.2 \pm2.1$ \\
$\Gamma(\eta_c)$[MeV] & $32.5 \pm 3.5$ & $32.5 \pm3.8$ \\
$\phi$ [rad] & $2.39 \pm 0.15$ & $4.34 \pm 0.09$ \\
$N^{\mathrm{obs}}$ & $757 \pm 31$ & $2033 \pm 61$ \\
$\mathcal{B}(\times10^{-3})$ & $0.90 \pm 0.04$ & $2.42 \pm 0.07$ \\
\hline\hline
\end{tabular}
\end{center}
\end{table}

\section{SYSTEMATIC UNCERTAINTY}\label{sec:sysU}
The systematic uncertainties in the BF measurements originate from several sources, as summarized in Table~\ref{table:sum_sys_Uncer}. They are introduced below.

\begin{enumerate}
\item \textit{Tracking}: The uncertainty of the $p/\bar{p}$ tracking is estimated by the control sample $J/\psi \to p\bar{p}\pi^+\pi^-$, and is assigned as $1.0\%$ per $p/\bar{p}$ track~\cite{Yuan:2015wga}.

\item \textit{PID}: The uncertainty of the PID of $p/\bar{p}$ hadrons is studied using the control sample $e^+e^-\to p\bar{p}\pi^0$, and is estimated to be $1.0\%$ per $p/\bar{p}$ track~\cite{BESIII:2012urf}.

\item \textit{Photon reconstruction}: The uncertainty of the photon selection is studied by using the control samples $J / \psi \rightarrow \pi^+\pi^-\pi^0$ and $e^{+} e^{-} \rightarrow \gamma \gamma$~\cite{photonrec}, and is determined to be $1.0\%$ per photon.

\item \textit{Kinematic fit}: The systematic uncertainty of the 4C kinematic fit is estimated to be $0.3\%$ by comparing the efficiencies with and without applying the helix correction~\cite{helix}, where the correction factors for protons are obtained by studying the control sample $\psipp \to p\bar{p}\pi^0$.

\item $\eta$ \textit{reconstruction}: The systematic uncertainty associated with $\eta$ meson reconstruction is dominated by the contribution from photon reconstruction, which has been addressed previously. The remaining uncertainty originates primarily from the difference in mass resolution between the data and the MC simulation due to the applied mass window selection. A comparison of the $M(\gamma\gamma)$ resolutions in data and MC simulation shows that this source of uncertainty is negligible.

\item \textit{Background vetoes}: To suppress backgrounds involving $\pi^0$ or $J/\psi$ mesons, mass windows are applied in the selection of data. A difference between the resolution in data and MC simulation can lead to a systematic uncertainty. The data–MC resolution difference is determined from the $\eta$ signal peak by fitting the data distribution with the signal MC simulation shape convolved with an additional Gaussian function.
The width of the Gaussian is taken as the extra smearing needed to account for the resolution difference. For each veto, the corresponding distribution in signal MC simulation is then smeared using this Gaussian resolution difference. The selection efficiency is determined with and without the smearing, and the difference is used to assign the associated systematic uncertainty. 

\item \textit{Physics model}: The difference in efficiency between the two signal MC samples, which are generated under the two physics models described in Sec.~\ref{Sec:BR_determined}, amounts to 1.1\%. This value is assigned as the corresponding systematic uncertainty.

\item \textit{Fit method}
    \begin{enumerate}
    \item \textit{Damping function}: The systematic uncertainty caused by the damping function is estimated by changing the damping function developed by KEDR~\cite{KEDR} with the one developed by CLEO~\cite{CLEO} ($f_d(E_{\gamma}) = \mathrm{exp}(-\frac{E_{\gamma}^2}{8\beta^2})$) in the fit, where the $\beta$ is a free parameter. 
    The resulting difference on the BF, $0.8\%$, is taken as the systematic uncertainty.
    \item \textit{Fit range}: In order to estimate the uncertainty from the fit range, we perform a Barlow test~\cite{Barlow:2002yb} of the significant deviation ($\zeta$) between the nominal fit and the systematic test. We vary the fit range of $M(p\bar{p}\eta)$ from $(2.780, 3.200)$ to $(2.804, 3.200)~\mathrm{GeV}/c^2$ with a step of $2~\mathrm{MeV}/c^2$. The relative systematic uncertainties are $1.3\%$ for the constructive-interference solution and $1.2\%$ for the destructive-interference solution, respectively.
    \item \textit{Efficiency curve}: The systematic uncertainty due to the efficiency curve is estimated by changing the line shape of a $2^{\rm{nd}}$-order Chebyshev function to the line shape of a $3^{\rm{rd}}$-order Chebyshev function in fitting the efficiency curve. The resulting difference in the BF is negligible.
    \item \textit{Shape of interference backgrounds}: The systematic uncertainty associated with the modeling of the interference background shape is evaluated using toy MC samples. These samples contain a virtual pseudo-scalar particle, implemented with a BW function characterized by a mass of $2.4~\mathrm{GeV}/c^2$ and a width of $150~\mathrm{MeV}$, and set to interfere coherently with the $\eta_c$ signal. The shape of this component in the $\eta_c$ mass region is then described with a polynomial function, as in the nominal fit. The difference between the true and fitted signal yields for the $\eta_c$ is taken as the corresponding systematic uncertainty, resulting in values of 1.6\% for the constructive-interference solution and 1.7\% for the destructive-interference solution. This study also indicates that, in the vicinity of the $\eta_c$ signal region, a polynomial function serves as a suitable approximation for describing a broad resonance located at a lower mass.  
    \item \textit{Fraction of interference component}: In the nominal fit, we assume that all processes yielding the final state $\gamma p\bar{p}\eta$ interfere fully with the $\eta_c$ signal, and that possible $p\bar{p}\eta$ systems having quantum numbers other than $0^{-+}$ would not interfere. To test this assumption, we introduce an additional non-interference component that also produces the same final state, thereby changing the scenario from full to partial interference. The shape of this added term is modeled by a second-order polynomial, with its amplitude floated and its shape parameters shared with those of the interference background component. The associated systematic uncertainty is taken as the difference between the results obtained with and without this extra term, yielding values of $18.4\%$ for the constructive-interference solution and $16.8\%$ for the destructive-interference solution.
    \item \textit{Number of non-interference backgrounds}: The systematic uncertainty associated with the normalization of the non-interference background component, $N_{\text{non}}$, is evaluated by varying its normalization factor to a value derived from the yields observed in the $\eta$ side band region. The resulting uncertainty is $8.9\%$, which applies equally to both the constructive- and destructive-interference solutions.
    \item \textit{Shape of non-interference backgrounds}: The inclusive MC simulation may not fully replicate the background present in the actual data. The systematic uncertainty associated with the modeling of the non-interference background shape is estimated conservatively by replacing the line shape of $f_{bg_2}$ with a $2^{\rm nd}$-order polynomial in the fit. The resulting uncertainties are 10.8\% for the constructive-interference solution and 3.0\% for the destructive-interference solution.
    \end{enumerate}
    
\item \textit{Quoted BFs}: According to PDG~\cite{ParticleDataGroup:2024cfk}, the uncertainties from BFs of $\psi(3686)\to \gamma \eta_c$ and $\eta \to \gamma \gamma$ decays are $13.9\%$ and $0.5\%$, respectively.

\item \textit{Number of $\psi(3686)$} events: The total number of $\psi(3686)$ events in data is determined to be $(2712.4\pm14.3)\times 10^6$~\cite{liucheng} from the data samples collected in 2009, 2012, and 2021. Therefore, this uncertainty is set to be $0.5\%$.
\end{enumerate}

The systematic uncertainties are summarized in Table~\ref{table:sum_sys_Uncer}, where the total uncertainties are obtained by adding all of them in quadrature, assuming all contributions are independent.

\begin{table}[!htbp]
\caption{\label{table:sum_sys_Uncer} Relative systematic uncertainties in the measurements of the BFs, where ``/'' indicates a negligible contribution. Here, ``Con.'' and ``Des.'' denote the constructive- and destructive-interference solutions, respectively. ``Total of Prod.'' is the total systematic uncertainty for the product BF of $\mathcal{B}[\psi(3686)\to \gamma\eta_c] \times \mathcal{B}[\eta_c\to p\bar{p}\eta]$, while ``Total of $\eta_c$'' corresponds solely to the uncertainty in $\mathcal{B}[\eta_c \to p \bar{p} \eta]$.}
\begin{tabular}{l l c c}
\hline\hline
\multirow{2}{*}{} & \multirow{2}{*}{Source} & \multicolumn{2}{c}{Sys. uncer.(\%)} \\
\cline{3-4}
 & & Con. & Des. \\
\hline
   & Tracking                                             & 2.0 & 2.0 \\
   & PID                                                  & 2.0 & 2.0 \\
   & Photon reconstruction                                & 3.0 & 3.0 \\
   & Kinematic fit                                        & 0.3 & 0.3 \\
   & $\eta$ reconstruction                                & /   & /   \\
   & $\pi^0 \rightarrow \gamma \gamma$ veto by $M(\gamma_1 \gamma_2)$ & 0.1 & 0.1 \\
   & $\pi^0 \rightarrow \gamma \gamma$ veto by $M(\gamma_1 \gamma_3)$ & 0.1 & 0.1 \\
   & $\pi^0 \rightarrow \gamma \gamma$ veto by $M(\gamma_2 \gamma_3)$ & 0.1 & 0.1 \\
   & $\psi(3686) \rightarrow \eta J / \psi$ veto by $RM(\gamma_2 \gamma_3)$ & 0.5 & 0.5 \\
   & Physics model                                        & 1.1 & 1.1 \\
\hline
\multirow{7}{*}{\rotatebox{90}{Fit method}} & Damping function                           & 0.8 & 0.8 \\
   & Fit range                                        & 1.3 & 1.2 \\
   & Efficiency curve                                     & /   & /   \\
   & Shape of interference backgrounds                    & 1.6 & 1.7 \\
   & Fraction of interfering component                    & 18.4 & 16.8 \\
   & Number of non-interference backgrounds               & 8.9 & 8.9 \\
   & Shape of non-interference backgrounds                & 10.8 & 3.0 \\
\hline
   & $\mathcal{B}(\eta \rightarrow \gamma \gamma)$        & 0.5 & 0.5 \\
   & Number of $\psi(3686)$ events                  & 0.5 & 0.5 \\
\hline
   & Total of Prod.                                       & 23.6 & 19.9 \\
\hline
   & $\mathcal{B}(\psi(3686) \rightarrow \gamma \eta_c)$  & 13.9 & 13.9 \\
\hline
   & Total of $\eta_c$                                    & 27.4 & 24.2 \\
\hline\hline
\end{tabular}
\end{table}

\section{SUMMARY}
Based on a sample of $(2712.4\pm14.3)\times 10^6$ $\psi(3686)$ events collected by BESIII in $2009$, $2012$, and $2021$, the $\eta_c \to p\bar{p}\eta$ decay is observed for the first time , via the $M\it1$ transition $\psi(3686)\to\gamma\eta_c$. The joint BF of the $\psi(3686)\to \gamma\eta_c$, $\eta_c\to p\bar{p}\eta$ decay chain is measured to be $(3.2 \pm 0.1 \pm 0.9)\times10^{-6}$ for the constructive-interference solution and $(8.7 \pm 0.3 \pm 2.1)\times10^{-6}$ for the destructive-interference solution, where the first uncertainties are statistical and the second systematic. The BF of $\eta_c\to p\bar{p}\eta$ is determined to be $\mathcal{B}(\eta_c\to p\bar{p}\eta)=(0.90 \pm 0.04 \pm 0.21 \pm 0.13)\times10^{-3}$ for the constructive-interference solution and $(2.42 \pm 0.07 \pm 0.48 \pm 0.34)\times10^{-3}$ for the destructive-interference solution, where the third uncertainties are due to the uncertainty in the BF of the $\psi(3686)\to \gamma \eta_c$ decay. 

Compared to theoretical predictions for $\mathcal{B}(\eta_c\to p\bar{p}\eta)$ discussed in the introduction, the measured results obtained in this work for both the constructive- and destructive-interference solutions are significantly higher than the NRQCD prediction by about one order of magnitude, while they are more consistent with the pQCD prediction within uncertainties.

\section{ACKNOWLEDGMENTS}
The BESIII Collaboration thanks the staff of BEPCII (https://cstr.cn/31109.02.BEPC) and the IHEP Computing Center for their strong support. This work is supported in part by the National Key R\&D Program of China under Contract Nos. 2025YFA1613900, 2023YFA1606000, 2023YFA1606704; the National Natural Science Foundation of China (NSFC) under Contract Nos. 12475197, 12275067, 11635010, 11935015, 11935016, 11935018, 12025502, 12035009, 12035013, 12061131003, 12192260, 12192261, 12192262, 12192263, 12192264, 12192265, 12221005, 12225509, 12235017, 12342502, 12361141819, and 12535005; the Chinese Academy of Sciences (CAS) Large-Scale Scientific Facility Program; the Strategic Priority Research Program of the Chinese Academy of Sciences under Contract No. XDA0480600; CAS under Contract No. YSBR-101; the 100 Talents Program of CAS; the Institute of Nuclear and Particle Physics (INPAC) and the Shanghai Key Laboratory for Particle Physics and Cosmology; ERC under Contract No. 758462; the German Research Foundation (DFG) under Contract No. FOR5327; Istituto Nazionale di Fisica Nucleare, Italy; the Knut and Alice Wallenberg Foundation under Contract Nos. 2021.0174, 2021.0299, and 2023.0315; the Ministry of Development of Turkey under Contract No. DPT2006K-120470; the National Research Foundation of Korea under Contract No. NRF-2022R1A2C1092335; the Natural Science Foundation of Henan Province under Contract No. 262300421060; the Science and Technology R\&D Program Joint Fund Project of Henan Province under Contract No. 225200810030; the Science and Technology Innovation Leading Talent Support Program of Henan Province under Contract No. 254200510039; the Henan Province Foreign Scientists Studio under Contract No. GZS2020032; the National Science and Technology Fund of Mongolia; the Polish National Science Centre under Contract No. 2024/53/B/ST2/00975; STFC (United Kingdom); the Swedish Research Council under Contract No. 2019.04595; and the U.S. Department of Energy under Contract No. DE-FG02-05ER41374.

\bibliographystyle{apsrev4-1}
\bibliography{References}

\end{document}